\documentclass[runningheads]{llncs}
\usepackage[T1]{fontenc}
\usepackage{sujitkc}
\usepackage{tabularx}

\begin{document}
\title{A Formal Verification Approach to Safeguard Controller Variables from Single Event Upset} 
\titlerunning{Safeguarding Controller Variables from SEUs}

%
%
\author{Ganesha\inst{1}\orcidID{0000-0001-9461-9231} \and
Sujit Kumar Chakrabarti\inst{2}\orcidID{0000-0001-8422-2900}}
\authorrunning{Ganesha and Sujit}
%
\institute{Samsung R\&D Institute India-Bangalore \and
International Institute of Information Technology, Bangalore, India
\email{\{ganesha, sujitkc\}@iiitb.ac.in}}
%


\maketitle              

\begin{abstract}
We present a method based on program analysis and formal verification to identify \emph{conditionally relevant variables} (CRVs) -- variables which could lead to violation of safety properties in control software when affected by \emph{single event upsets} (SEUs). Traditional static analysis can distinguish between relevant and irrelevant variables. However, it would fail to take into account the conditions specific to the control software in question. This can lead to false positives. Our algorithm employs formal verification to avoid false positives. We have conducted experiments that demonstrate that CRVs indeed are fewer in number than what traditional static analysis can detect and that our algorithm is able to identify this fact. The information provided by our algorithm could prove helpful to a compiler while it does register allocation during the compilation of the control software. In turn, this could cause significant reduction in the cost of controller chips. 

\keywords{Program Analysis,
CRV (Conditional Relevance of a Variable),
SEU (Single Event Upset),
Hardening,
Model Checking,
Formal Verification,
Program Slicing}

\end{abstract}

\section{Introduction}
\subsection{Single Event Upsets}
Control systems are typically implemented as control algorithms running on a microcontroller or \emph{system-on-chip} (SoC) devices. Often, such embedded controllers function in harsh environmental conditions. In such cases, the computing components get exposed to a rare random phenomenon called \emph{single event upset} (SEU) \cite{ref_seu1}. When an SEU happens, a bit in the chip would flip its value. This could lead the ongoing computation to fall off track leading to errors -- often with catastrophic consequences in safety critical applications, e.g. automotive, industrial automation, aerosphere, oil and gas, mining, healthcare etc. 

\subsection{Hardening}
A common technique used in the industry to safeguard electronic components from SEUs is called \emph{hardening} \cite {ref_url2}. The hardened part of the silicon can withstand the causes of SEUs, and bit-flips are avoided. Hardening causes the cost of the chip to go up. Hence, it would be better to harden only a portion of the chip that balances the risk of SEUs with the rise in manufacturing costs of the chip. However, this would require that code running on the chip to be generated with this fact taken  into account. During runtime, the parts of the code that must be protected from SEUs must reside in the hardened part of the chip while the other `non-critical' components of computation can reside in the non-hardened part of the chip. In this paper, we present a method to statically identify variables which are `critical' in the sense that their getting affected by SEUs would likely throw the computation off-track leading to unacceptable errors.

\subsection{Our Contribution}
In this paper, we present an approach for dividing the variables of the control algorithm into two categories: \emph{conditionally relevant variables} (CRVs) -- those which should be placed in the hardened part of the silicon, and the other  -- called \emph{conditionally irrelevant} or non-CRVs -- that can be left unguarded by being placed in the unhardened part. We note that discovering the precise set of CRVs is undecidable. Static analysis techniques, e.g., slicing, can be used to detect CRVs, however they give a sound but overapproximate set of CRVs. Our algorithm uses formal verification (model checking) that helps us tighten this bound without compromising soundness. 

This paper makes the following contributions:
\begin{enumerate}
\item We introduce and formally define the idea of \emph{conditional relevance} of a variable.
\item We use a combination of static analysis (program slicing) and formal verification (software model checking) to identify the conditional relevance/irrelevance of a variable.
\item We present experimental evaluation of our approach on a number of example programs and demonstrate the fact that: 
	\begin{enumerate}
	\item There indeed are present conditionally irrelevant variables in control softwares.
	\item Using formal verification over and above traditional static analysis (i.e. static program slicing), helps discover additional conditionally irrelevant variables, avoiding false positives and over-conservative decisions. 
	\end{enumerate}
\end{enumerate}

\begin{figure}
\begin{center}

\begin{tikzpicture}
\node[](prog){Program};
\node[rectangle, draw, thick, rounded corners, right=of prog](analysis){Analysis};
\node[rectangle, align=center, right=of analysis](cp){Critical \\ Part};
\node[rectangle, draw, thick, rounded corners, right=of cp](comp){Compiler};
\node [draw, thick, rectangle split, align=center, rectangle split parts=2, rectangle split part fill={Gray!50,none}, right=of comp] (output)
{
	Hardened
    \nodepart{two} Not \\ hardened
};

\draw[->, Red, thick](prog) -- ++ (0:1cm) -- ++ (90:1cm) -| (comp);
\draw[->, Red, thick](prog) -- (analysis);
\draw[->, Red, thick](analysis) -- (cp);
\draw[->, Red, thick](cp) -- (comp);
\draw[->, Red, thick](comp) -- ++ (0:1cm) |-(output.150);
\draw[->, Red, thick](comp) -- ++ (0:1cm) |-(output.200);
\end{tikzpicture}
\end{center}
\caption{Hardening}
\label{f:Harden}
\end{figure}
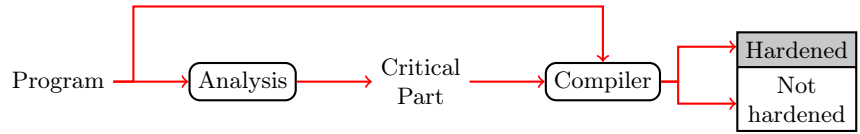

We believe that the method we present in this paper can help bring down the manufacturing cost of computing chips, especially in control software domain, by providing useful inputs to the compiler. Figure~\ref{f:Harden} illustrates the selective hardening process, where a program undergoes \emph{analysis} (i.e. using the algorithm presented in this paper) to identify the critical parts which cannot be allowed to suffer SEUs. The information about these parts are passed on to the compiler which generates code to ensure that variables reside in the hardened and non-hardened parts of the chip during runtime based on their criticality.

\section{Motivating Example}

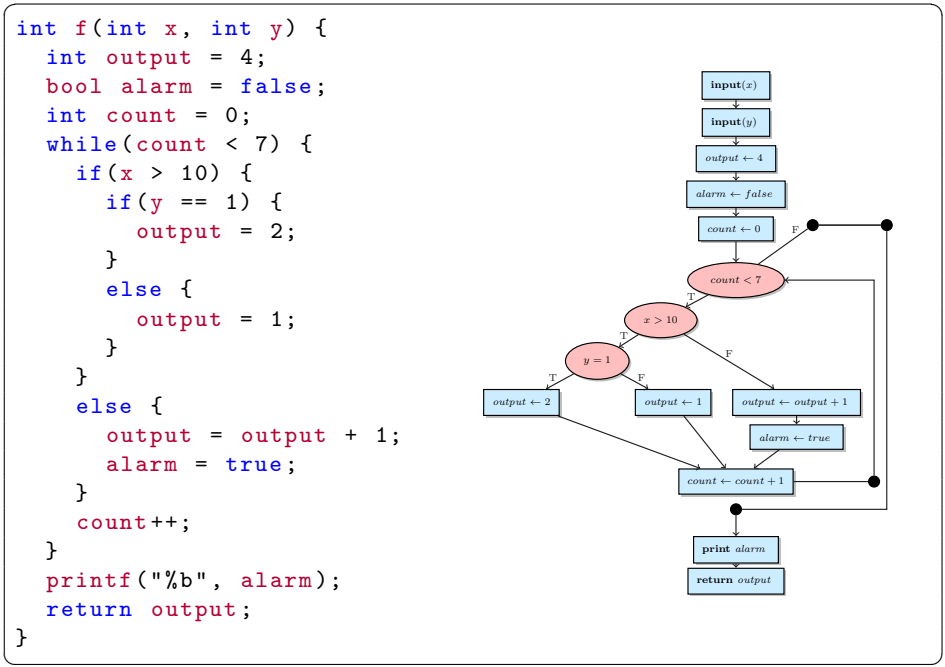
\begin{figure}
\begin{center}
\begin{tabular}{c @{\hspace{1cm}} c}
\begin{lstlisting}[style=jc]
int f(int x, int y) {
  int output = 4;
  bool alarm = false;
  int count = 0;
  while(count < 7) {
    if(x > 10) {
      if(y == 1) {
        output = 2;
      }
      else {
        output = 1;
      }
    }
    else {
      output = output + 1;
      alarm = true;
    }
    count++;
  }
  printf("%b", alarm);
  return output;
}
\end{lstlisting}
&

\begin{minipage}{0.65\textwidth}
\begin{scriptsize}
\resizebox{0.7\textwidth}{!}{
\begin{tikzpicture}
\node[bb](b1a){
$\textbf{input}(x)$
};
\node[bb, below=0.2cm of b1a](b1b){
$\textbf{input}(y)$
};
\node[bb, below=0.2cm of b1b](b1c){
$\mathit{output} \gets 4$ 
};
\node[bb, below=0.2cm of b1c](b1d){
$\mathit{alarm} \gets false$
};
\node[bb, below=0.2cm of b1d](b1e){
$count \gets 0$
};
\node[db, below=0.5cm of b1e](d2){$count < 7$};
\node[db, below left=0.5cm of d2](d3){$x > 10$};
\node[db, below left=0.5cm of d3](d4){$y  = 1$};
\node[bb, below left=0.5cm of d4](b5){$\mathit{output} \gets 2$};
\node[bb, below right=0.5cm of d4](b6){$\mathit{output} \gets 1$};
\node[bb, right=0.5cm of b6](b7a){
$\mathit{output} \gets \mathit{output} + 1$
};
\node[bb, below=0.2cm of b7a](b7b){
$\mathit{alarm} \gets true$
};

\node[bb, below=5.5cm of b1e](b8){$count \gets count + 1$};
\node[bb, below=1cm of b8](b9a){
$\textbf{print}\ \mathit{alarm}$
};
\node[bb, below=0.1cm of b9a](b9b){
$\textbf{return}\ \mathit{output}$
};
\draw[->, Black, thick] (b1a) to (b1b);
\draw[->, Black, thick] (b1b) to (b1c);
\draw[->, Black, thick] (b1c) to (b1d);
\draw[->, Black, thick] (b1d) to (b1e);
\draw[->, Black, thick] (b1e) to (d2);
\draw[->, Black, thick] (d2) to node[above, near end]{T}(d3);
\draw[->, Black, thick] (d3) to node[above, near end]{T} (d4);
\draw[->, Black, thick] (d4) to node[above, near end]{T} (b5);
\draw[->, Black, thick] (d4) to node[above, near end]{F} (b6);
\draw[->, Black, thick] (d3) to node[above]{F} (b7a);
\draw[->, Black, thick] (b5) to (b8);
\draw[->, Black, thick] (b6) to (b8);
\draw[->, Black, thick] (b7a) to (b7b);
\draw[->, Black, thick] (b7b) to (b8);

\node[jn, right=1.8cm of b8] (j1){};
\draw[-, Black, thick] (b8) to (j1);
\draw[->, Black, thick] (j1) |- (d2);

\node[jn, above right=1.3cm of d2] (j2){};
\node[jn, right=1.5cm of j2] (j3){};
\node[jn, above=0.5cm of b9a] (j4){};
\draw[-, Black, thick] (d2) to node[above, near end]{F}(j2);
\draw[-, Black, thick] (j2) to (j3);
\draw[-, Black, thick] (j2) to (j3);
\draw[-, Black, thick] (j3) |- (j4);
\draw[->, Black, thick] (j4) to (b9a);
\draw[->, Black, thick] (b9a) to (b9b);
\end{tikzpicture}

}
\end{scriptsize}
\end{minipage}
\end{tabular}
\end{center}
\caption{Motivating Example}
\label{f:mot}
\end{figure}

In Figure~\ref{f:mot}, we show a fragment of code on the left side and its control flow graph on the right side. The program computes and returns the value of the output variable $output$. Suppose that there is correctness condition imposed on the system that says that $output$ should be less than or equal to 10 before it is returned. We can see that it is possible to violate this condition in the given code. The program iterates through the outer while loop exactly 7 times. If the branch condition ($x > 10$) is false, $output$ would be incremented by 1 in each iteration. In 7 iteration, its value would be 11, which would violate the safety property. But, an interesting aspect of this situation is that the above computation never uses the value of $y$. This is not to say that $output$ is not dependent on $y$; in fact, it \emph{is} dependent on $y$. A traditional static slicing would reveal this to us. But this technique would fail to discover that $output$ doesn't depend on $y$'s value when it violates the safety property. This is to say that $y$ is not relevant as far as the safety property $output \le 10$ is concerned. We use formal verification to identify such cases.

\section{Terms and Concepts}

\subsection{Dependencies and Static Program Slicing}

In a  program, when one part, say $A$, affects the computation happening at some other part, say $B$, there is said to be a dependency between $A$ and $B$, or $B$ is dependent on $A$. There are two types of dependencies: \emph{data dependency} and \emph{control dependency}. When the computation at $A$ is used as input for $B$, $B$ is said to be \emph{data dependent} on $A$. When the execution of $B$ depends on the computation at $A$, $B$ is said to be \emph{control dependent} on $A$.

A \emph{program slice} (or \emph{slice} for short) in a program is the part of the program related in some causal way to a specific program location. For example, a backward slice gives those parts of the program which, during some run, influence the value of a variable at a program point of interest. This pair of variable ($v$) and program point ($l$) is known as the \emph{slicing criterion}. Program slicing includes a range of techniques in program analysis to discover slices, which can be static, dynamic, backward, or forward. Slicing has been successfully and widely used to solve many software engineering problems.

\subsection{Observations}
Static slicing yields sound results, i.e. it will not miss identifying any CRV. However, it will often yield imprecise results. For example, in the Figure~\ref{f:mot}, it will not be able to detect the fact that $y$ is not a CRV for $output > 10$ at the return statement. In this work, we use program slicing  and formal verification to identify the conditionally relevant variables. As a result, the compiler can be guided to place conditionally relevant variables in the hardened registers during register allocation. These bounds will often be tighter than the ones obtainable through traditional static analysis. This will make it possible to harden a smaller part of the chip, leading to cost saving without loss of safety guarantee.

\subsection{The safety property $\Phi$}
A safety property defines the safety of the computation and is often a weaker condition than strict correctness. This is because, in close-loop feedback control systems, many glitches (e.g. those arising out of SEUs) do not persist beyond a control cycle and are often self-correcting due to the dynamic properties of the controlled plant. In other words, the plant often acts like a low-pass filter that removes high frequency transients before they can adversely affect the system function.

As an example, suppose that in a temperature control system, if the value of the variable $o$ corresponding to the output heating rate gets affected by SEU, the plant temperature $\theta$ is not going to change very fast due to the high time constants associated with heating. As a result the next value of the same variable $o$, computed based on the error value ($\theta_{ref} - \theta$), will not be significantly influenced by the fact that it had got disturbed due to an SEU in the last control cycle. The system is capable of `forgetting' one-time upsets like this in many cases. Any exception (either internal or environmental) causing value changes in $o$ would cause \emph{unacceptable} changes in the controlled plant only if it persists for some time. The safety property formulates the criterion that defines an acceptable or safe behaviour from the point-of-view of the controlled plant.  

More formally, the safety property is a function that defines if the output computed at the \emph{output point}, i.e. end of each control cycle, is `safe'. This is derived from knowledge about the physical properties of the plant being controlled. For now, this process is based on domain knowledge.

For example, it may be stipulated that the computed control output $o$ must not be outside a certain range, say $(r_{min}, r_{max})$ for more than 5 consecutive times. Suppose $O$ is a buffer which is used to store the last 5 values of the output variable, with $O_4$ being the last computed value and $O_0$ being the fifth most recent value of $o$. Then,

\begin{equation*}
\Phi(O) = \neg( \bigwedge\limits_{i=0}^4(O_i < r_{min} \lor O_i > r_{max}))
\end{equation*}

\subsection{Conditionally Relevant Variable (CRV)}
We now present the idea of conditionally relevant variables by first giving an informal explanation and then presenting a formal definition.

\emph{Relevant variables} are those on which the output of the computation is dependent. Static analysis, e.g. program slicing, can identify relevant variables. On the other hand, a \emph{conditionally} relevant variable (CRV) is one which affects the satisfaction of a given safety property $\Phi$. Conversely, a \emph{conditionally irrelevant variables} (non-CRV) is one whose value does not influence the satisfaction of the given safety property. Therefore, a variable may be relevant, but may be conditionally irrelevant as per the given safety property.

More specifically, a conditionally relevant variable is one whose value -- if changed randomly during program execution -- would lead to a change in the value of the safety property in at least one possible execution.

Now, we present a more formal definition of conditional relevance.

Suppose that $V$ is the set of all variables in the controller program. Let $I (\subseteq V) = \{x_1, x_2, .., x_n\}$ be the set of all input variables of the program, i.e., their values are set through interactions with the environment, e.g. controlled plant, human operator/user, sensors or other subsystems. Let's say that, for a given run of the program, the input values received by an input variable $x_i$ are $\mathcal{I}(x_i) = \{x_{i,1}, x_{i,2}, ...\}$. We define an input vector as a sequence of all values assigned to all input variables: $\mathcal{I} = \bigcup\limits_{x_i \in I} \mathcal{I}(x_i)$. We denote the set of all possible input vectors by $\mathbf{I}$.

A trace is the sequence of program locations (possibly with multiple occurances due to loops) visited during an execution of the program. In absence of any upsets, each input vector would map to a unique execution trace, given by $\pi(\mathcal{I}) = \{l_1, l_2, ...\}$ during the execution corresponding to $\mathcal{I}$. The output location is the program point $p$ that immediately precedes the command that causes the result variable to be output by the controller in each control cycle. An output point set $P(\mathcal{I}) \subseteq \pi(\mathcal{I})$ for an input vector $\mathcal{I}$ is the set of occurances of the output location $p$ in $\pi(\mathcal{I})$.

Let $l \in \pi(\mathcal{I})$ be the location where a variable $x \in V$ suffers an SEU. As a result, the trace that evolves is $\pi'(\mathcal{I})$. We note that this may cause $\mathcal{I}$ to no more be completely input to the program due to the modified trace possibly diverging away from the original. However, the prefix of the modified input vector $\mathcal{I}'$ upto $l$ would be identical to that of $\mathcal{I}$, and that is what we are interested in. Correspondingly, the output point set thus visited is $P'(\mathcal{I}) \subseteq \pi'(\mathcal{I})$.

Formally, the conditional relevance of a variable $x$ is defined as:

 \begin{equation}
\begin{split}
CRV(x) &= \exists\ \mathcal{I} \in \mathbf{I} \text{ such that} \\
       &(\forall p \in P(\mathcal{I})\ |\ \Phi_p = true \land \exists p' \in P'(\mathcal{I})\ |\ \Phi_p' = false) \\
       &\text{\hspace{3cm}}\lor \\
       &(\forall p' \in P'(\mathcal{I})\ |\ \Phi_p' = true \land \exists p \in P(\mathcal{I})\ |\ \Phi_p = false)
\end{split}
\end{equation}

In other words,
\begin{enumerate}
\item Conditional relevance requires the safety property to be violated for at least one input vector when $x$ is affected by an SEU  ($\Phi_p' = false$) in the condition that the safety property is never violated for that input vector in absence of an SEU affecting $x$ ($\Phi_p = true$).
\item Also, conditional relevance requires the safety property to hold true in presence of SEU on $x$ while there is violation in its absence.
\end{enumerate}

When an SEU affects $x$, the first condition above leads to an occurance of fault, while the second case could cause a fault getting masked due an SEU. Both these situations are undesirable. Please note that the second case is likely to be rarer, but possible.

All variables $x \in V$ for which $CRV(x) = true$ are conditionally relevant variables. All other variables are conditionally irrelevant variables.

\section{Our Approach}
\subsection{Overall Architecture}

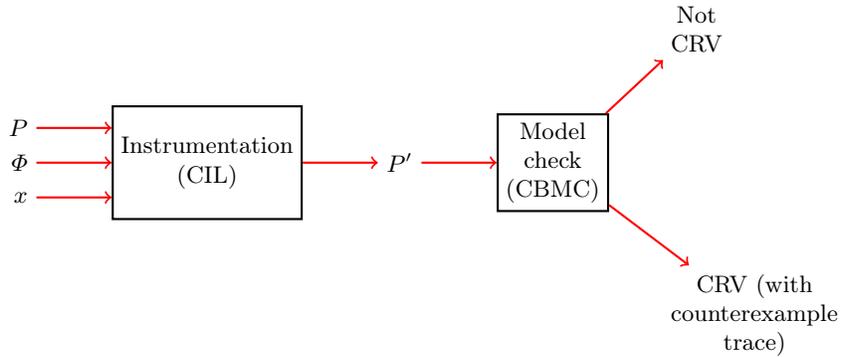
\begin{figure}
\begin{center}
\begin{tikzpicture}
\node[rectangle, draw, thick, align=center, minimum height=1.5cm](instr) {Instrumentation \\ (CIL)};
\node[left=1cm of instr.160](p) {$P$};
\node[left=1cm of instr](phi) {$\Phi$};
\node[left=1cm of instr.200](x) {$x$};
\node[right=of instr](pdash) {$P'$};
\node[rectangle, draw, thick, align=center, right=of pdash](mc) {Model \\ check \\ (CBMC)};
\node[align=center, above right=1cm of mc](r1) {Not \\ CRV};
\node[align=center, below right=1cm of mc](r2) {CRV (with \\ counterexample \\ trace)};

\draw[->, Red, thick] (p) -- (instr.160);
\draw[->, Red, thick] (phi) -- (instr);
\draw[->, Red, thick] (x) -- (instr.200);
\draw[->, Red, thick] (instr) -- (pdash);
\draw[->, Red, thick] (pdash) -- (mc);
\draw[->, Red, thick] (mc) -- (r1);
\draw[->, Red, thick] (mc) -- (r2);

\end{tikzpicture}
\end{center}
\caption{Overall approach}
\label{f:arch}
\end{figure}

Figure~\ref{f:arch} shows our overall approach. $P$ is the control program that is being analysed, $\Phi$ is the safety property and $x$ is the \emph{variable under investigation}, i.e. we are investigating if $x$ is a CRV or not. The first step is to instrument $P$ and generate the instrumented version $P'$. This is given as an input to a model checker. The model checker either establishes that $x$ is not a CRV, or establishes that it is a CRV, along with a counterexample trace through $P'$ that shows how an SEU affecting $x$ may lead to the violation of the safety property.

\subsection{Instrumentation} \label{s:instr}

Our overall approach involves instrumenting the control procedure $P$ in preparation for verification. We call the resultant instrumented procedure $P'$. Consider the structure of $P$ as shown in Algorithm~\ref{a:instru-algo}. $I$ is the initialisation part of the program. $B$ is the part of the main loop which computes the output variable $o$.
The sequence $S$ is created by taking $B$ and sequencing with it \lstinline[style=jc]|$O$.append($o$)| and \lstinline[style=jc]|$\phi$ := $\Phi(O)$|. As $P'$ is created by merging the original and modified version of $P$, we ensure to replace all variables in $P$ with another variable, in both $I$ and $S$, denoted by $I'$ $\gets$ $I[V'/V]$ and $S'$ $\gets$ $S[V'/V]$ respectively. Here, $V$ is set of variables in $P$ and $V'$ is the set of new variables. $S'$ is derived by inserting \(\text{\lstinline[style=jc]|mimic_seu_effect(&x)|}\)  before every use of $x \in V$, mimicing SEUs. $O$ is a buffer that is used to maintain the last $n$ values of the output variable $o$. \lstinline[style=jc]|$O$.append($o$)| results in $o$ being added as the last element of $O$. If $O$ already had $n$ elements, its first element is dropped in order to make space for the latest value of $o$. A variable under investigation, say $x$, may suffer from an SEU anytime during the execution of the program. Hence, we insert a \(\text{\lstinline[style=jc]|mimic_seu_effect(&x)|}\) command (see section~\ref{f:simseu}) before every use of $x$.

\begin{algorithm}

\begin{algorithmic}
\Function{instrument}{$P$}

\Comment{$P$ is a control algorithm with the following structure:

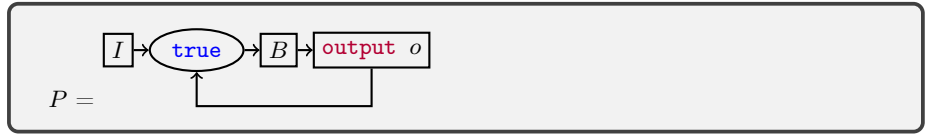
\begin{tcolorbox}[width=\textwidth]
$P$ =
\begin{tikzpicture}
\node[rectangle, draw, thick](I){$I$};
\node[ellipse, draw, thick, right=0.2cm of I](d1){\lstinline[style=jc]|true|};
\node[rectangle, draw, thick, right=0.2cm of d1](B){$B$};
\node[rectangle, draw, thick, right=0.2cm of B](r){\lstinline[style=jc]|output\ $o$|};
\coordinate (j1) at ($(r.south)+(-90:0.5cm)$);

\draw[->, thick](I) -- (d1);
\draw[->, thick](d1) -- (B);
\draw[->, thick](B) -- (r);
\draw[->, thick](r) -- (j1) -| (d1);
\end{tikzpicture}
\end{tcolorbox}
}

\Comment where $B$ is the body of an arbitrary control computation. \\

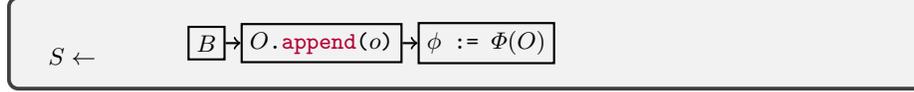
\State
\begin{tcolorbox}[width=\textwidth]
$S \gets$ \hspace{1cm}
\begin{tikzpicture}
\node[rectangle, draw, thick](B){$B$};
\node[rectangle, draw, thick, right=0.2cm of B](ap){\lstinline[style=jc]|$O$.append($o$)|};
\node[rectangle, draw, thick, right=0.2cm of ap](ph1){\lstinline[style=jc]|$\phi$ := $\Phi(O)$|};

\draw[->, thick](B) -- (ap);
\draw[->, thick](ap) -- (ph1);
\end{tikzpicture}
\end{tcolorbox}

\State $I'$ $\gets$ $I[V'/V]$
\State $S'$ $\gets$ $S[V'/V]$ \\

\State
\begin{tcolorbox}[width=\textwidth]
\textbf{return} 
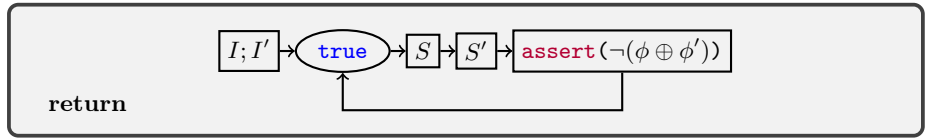
\hspace{1cm}
\begin{tikzpicture}
\node[rectangle, draw, thick](I){$I; I'$};
\node[ellipse, draw, thick, right=0.2cm of I](d1){\lstinline[style=jc]|true|};
\node[rectangle, draw, thick, align=left, right=0.2cm of d1](S){$S$};
\node[rectangle, draw, thick, align=left, right=0.2cm of S](Sdash){$S'$};

\node[rectangle, draw, thick, right=0.2cm of Sdash](as){\lstinline[style=jc]|assert($\neg(\phi \oplus \phi')$)|};
\coordinate (j1) at ($(r.south)+(-90:0.5cm)$);

\draw[->, thick](I) -- (d1);
\draw[->, thick](d1) -- (S);
\draw[->, thick](S) -- (Sdash);
\draw[->, thick](Sdash) -- (as);
\coordinate (j1) at ($(as.south)+(-90:0.5cm)$);
\draw[->, thick](as) -- (j1) -| (d1);
\end{tikzpicture}
\end{tcolorbox}
\EndFunction

\end{algorithmic}
\caption{Instrumentation algorithm}
\label{a:instru-algo}
\end{algorithm}

\subsection{Mimicking SEU Effects} \label{s:seu_sim} 
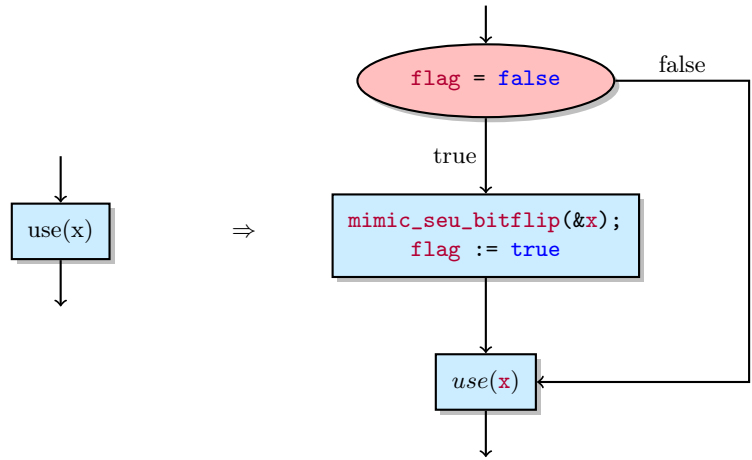
\begin{figure}[h!]
\begin{center}
\begin{tabular}{c @{\hspace{0.5cm}$\Rightarrow$\hspace{1cm}}c}
\begin{minipage}{0.2\textwidth}
\begin{tikzpicture}
\node[bb] (usex) {use(x)};

\draw[<-, thick](usex) -- ++(90: 1);
\draw[->, thick](usex) -- ++(-90: 1);
\end{tikzpicture}
\end{minipage}
&
\begin{minipage}{0.6\textwidth}
\begin{tikzpicture}
\node[db] (init) {\lstinline[style=jc]|flag = false|};
\node [bb, align=center, below=of init] (seu) {\lstinline[style=jc]|mimic_seu_bitflip(&x);| \\ \lstinline[style=jc]|flag := true|};
\node[bb] (usex) [below=of seu] {$use$(\lstinline[style=jc]|x|)};

\draw[->, thick] (init) -- node[left]{true}(seu);
\draw[->, thick] (seu) -- (usex);

\draw[->, thick]
  (init) -- node[above]{false} ++ (3.5,0)      
              |- (usex);      

\draw[<-, thick](init) -- ++(90: 1);
\draw[->, thick](usex) -- ++(-90: 1);

\end{tikzpicture}
\end{minipage}

\end{tabular}
\end{center}
\caption{Method for Mimicking SEU Effects}
\label{f:SS}
\end{figure}

To mimic SEU, we inject bit flip on the variable under investigation before each of its use.  This is done through code instrumentation. We have developed a tool that is built upon C Intermediate Language (CIL)~\cite{ref_cil} that does the instrumentation automatically. This inserts calls to the helper function \(\text{\lstinline[style=jc]|mimic_seu_bitflip|}\) into the instrumented program. A simplified version of the logic of inserting bit-flip is shown in Figure~\ref{f:simseu}.  SEU can happen at any place of the program. We also ensure that the SEU can happen up to only once (using \lstinline[style=jc]|flag| variable).  The method for mimicking SEU effects  is explained in Figure~\ref{f:SS}. 

Figure~\ref{f:simseu}  shows the code for injecting a Single Event Upset (SEU) into a target variable. A single-bit flip is introduced at a random bit position using XOR logic. Instrumentation preserves original semantics of the program apart from the calls to the function \(\text{\lstinline[style=jc]|mimic_seu_effect|}\). 
We call \(\text{\lstinline[style=jc]|mimic_seu_effect|}\)
before every use of \lstinline[style=jc]|x|, where \lstinline[style=jc]|x| is the variable under investigation. The function internally uses \(\text{\lstinline[style=jc]|nondet_int()|}\) from CBMC to mimic random bit flip.

\begin{figure}
\begin{lstlisting}[style=jc]
// Function to generate a nondeterministic integer within the
// range [1, 32]
int nondet_int_range_1_32() {
    int value = nondet_int() % 32 + 1;
    return value;
}

int mimic_seu_bitflip(int value, int bit_pos) {
    int mask = 1 << bit_pos;
    return (value ^ mask); // XOR operation for bit flip
}

// Ensures that an SEU is introduced only once for the
// variable under investigation
void mimic_seu_effect(int *invest_var) {
    static int count = 0;
    if(count == 0) {
        int bit_pos = nondet_int_range_1_32();
        *invest_var = mimic_seu_bitflip(*invest_var, bit_pos);
        count++;
    }
}
\end{lstlisting}
\caption{Code to Mimic SEU}
\label{f:simseu}
\end{figure}

\section{Experimental Results}

\subsection{Experimental Setup}
The main goal of our experiment is to validate and demonstrate the effectiveness of our proposed algorithm. 
We use a set of controller algorithms written in C language as case studies, as shown in Table~\ref{tab:crv_comparison}. We have used Frama-C version 30.0~\cite{ref_framac} for program slicing and CBMC version 5.12  for model checking. The slicing criterion for each case study was selected based on a safety property defined in terms of the controller output variable at output point (i.e. the program location just before when output value is written to actuator/plant) of the controller procedure.

We first slice the input program to remove everything except the relevant program statements w.r.t. to the safety property. The sliced input code is fed into an instrumentation step (see Section~\ref{s:instr}). Here, we apply our SEU mimicking method (Section~\ref{s:seu_sim}) to the variable under investigation and create the instrumented version of the program. The instrumented code is then fed to a model checker (CBMC), along with the safety property. CBMC applies bounded model checking and determines whether the safety property is violated or not. If it is violated, CBMC provides a counterexample trace.

\subsection{Case Studies and Evaluation}
To evaluate the effectiveness of our algorithm we have tested it on one controller algorithms (\emph{Motivating Example}) which mimics the real world closed-loop systems. Additionally, we have verified our algorithm on two programs (\emph{Temperature Control} and \emph{Fan Speed Control}) not having a structure similar to a typical control algorithm (i.e. with an outer while loop and control output being computed in the loop body) to demonstrate its general applicability. We have compared our algorithm with static slicing in terms its precision in detecting CRVs. We consider below points during our evaluation.

\begin{enumerate}
\item Total number of variables examined
\item CRVs identified in both the methods
\item Non-CRVs retained by static slicing but correctly eliminated by our approach
\item The safety property used for each program
\end{enumerate}

\begin{table}[h]
\centering
\begin{tabular}{|>{\centering\arraybackslash}p{2.0cm}|>{\centering\arraybackslash}p{1.2cm}|>{\centering\arraybackslash}p{0.9cm}|>{\centering\arraybackslash}p{0.9cm}|>{\centering\arraybackslash}p{0.9cm}|>{\centering\arraybackslash}p{1.2cm}|>{\centering\arraybackslash}p{3.0cm}|}
\hline
\textbf{Program Name} & \textbf{LoC} & \textbf{T} & \textbf{S} & \textbf{M} & $\boldsymbol{\eta}$ (\%) & $\boldsymbol{\Phi}$ \\ \hline
Motivating Example & 30 & 5 & 4 & 1 & 25\% & $output \leq 10$ \\ \hline
Temperature Control & 45 & 6 & 4 & 2 & 50\% & $temperature \leq 30$ \\ \hline
Fan Speed Control & 45 & 5 & 4 & 1 & 25\% & $fan\_speed \leq 100$ \\ \hline
\end{tabular}
\caption{Comparison of CRV Detection: Static Slicing vs Our Approach}
\label{tab:crv_comparison}
\end{table}

As shown in Table~\ref{tab:crv_comparison}, some of the variables identified as relevant variables (RVs) by static slicing are found to be conditionally irrelevant (flagging them as non-CRVs) by our algorithm. For instance, in each of the \emph{Motivating Example} and \emph{Fan Speed Control} programs, our approach eliminates one such variable that static slicing retains. In the \emph{Temperature Control} case, two additional non-CRVs were correctly discarded by our method. These results support our claim that the proposed approach improves over static slicing by reducing false positives and more precisely identifying the conditionally relevant variables.

\vspace{0.2cm}
\noindent
We define the following parameters for the results shown in Table~\ref{tab:crv_comparison}. $T$ represents the total number of input variables in the program. $S$ denotes the number of relevant variables retained by static slicing (i.e., RVs). $M$ represents the number of non-CRVs detected and eliminated by our approach. $\eta$ denotes the efficiency of our method, and $\Phi$ represents the safety property being verified for each benchmark. LoC indicates the number of lines of code in the program.

\vspace{0.2cm}
\noindent
We calculate the efficiency $\eta$, as $\eta = \frac{M}{S} \times 100$, where $S$ is the number of relevant variables retained by static slicing and $M$ is the number of non-CRVs correctly eliminated by our approach. A higher value of $\eta$ indicates a greater number of false positives successfully removed by our method, demonstrating its advantage over slicing. For example, if static slicing retains $S=4$ variables and our algorithm eliminates $M=1$ non-CRV, the efficiency is calculated as $\eta = \frac{1}{4} \times 100 = 25\%$.

\section{Related Work}

\subsection {Program Slicing and Its Limitations}
Program slicing is well known method of program analysis and debugging. Traditional program slicing, such as those introduced by Weiser \cite{ref_proc1} and surveyed by Tip \cite{ref_proc2}, focuses on semantic dependencies to slice the program. However these approaches often lead to over approximation, i.e. they include the variables that semantically do not impact the program behaviour. Recent studies in program slicing explored some enhancements to existing slicing techniques. For example,  \cite{ref_proc3} proposes predictive slicing using machine learning to anticipate runtime behaviour and \cite{ref_proc4} introduces field-sensitive slicing to improve precision. Despite these enhancements to program slicing, the actual challenge, i.e. of identifying actual set of variables that are really critical for program correctness, especially under controller safety propertys,  still remains the same.

\subsection {Software Model Checking and Verification Tools}
Today, state-of-the-art software model checking is significantly evolved. Tools such as CBMC \cite{ref_cbmc1} and ESBMC \cite{ref_cbmc2} that can formally verify C and C++ programs. These tools are used to identify bugs and to verify the program correctness. Recently, CBMC added a symbolic shadow memory as an additional mechanism to verify memory safety \cite{ref_cbmc3}. Our work builds on top of software model checking to detect CRVs in controller algorithms.

\subsection {SEU Mitigation Strategies}
Since SEUs occur in control chips, they impact system reliability  especially in the harsh environmental condition.  Conventional mitigation approaches are concentrated on  hardware level solutions such as error correcting codes and redundancy. \cite{ref_seu_mitigation1} proposes a self-adaptive SEU mitigation scheme based on ECC and refreshing technique for embedded systems. \cite{ref_swift} proposes a software-only fault tolerance technique, SWIFT, to detect and recover from transient faults such as SEUs and Single Event Transients (SETs) in microprocessors. Our work introduces a novel approach by leveraging static analysis and formal methods to identify CRVs. 

\section{Conclusion}
In this paper, we have introduced the concept of conditionally relevant variables (CRVs) -- variables whose values must not be affected by random changes at runtime to ensure safe execution of the control program. This is represented by the satisfaction of a safety property. Our algorithm determines the conditional relevance of variables in a control software using static analysis and formal verification. We have demonstrated that the set of CRVs is often a proper subset of all relevant variables. This fact could be used to provide more fine-grained control to the compiler so as to ensure that only CRVs reside in the hardened portions of the silicon. This would make the idea of partial hardening (i.e. hardening only portions of the chip) practical without compromising safety or performance. This, in turn, would help bring down chip manufacturing costs.
\subsection{Future Work}

Currently, we have restricted our attention to only variables in terms of their conditional relevance. The idea can be extended to the verification of conditional relevance of program locations too. Further, in its current form, our algorithm can analyse single-procedure control programs. In future we will extend our research to handle multi-procedure programs using inter-procedural analysis. Also, we plan to explore other model checking tools beyond CBMC to ensure soundness over the complete state space. Above enhancements will enable us to apply our algorithm to industrial-strength case studies. Finally, we aim to automate the formulation of safety properties based on system specification.

%
%
%

\end{document}